\documentclass[
  aps,
  prl,
  twocolumn,
  reprint,
  nofootinbib,
  preprintnumbers,
  superscriptaddress,
  amsmath,
  amssymb
]{revtex4-2}
\usepackage{graphicx}
\usepackage{bm}
\usepackage{amssymb}
\usepackage{amsmath}
\usepackage{comment}
\usepackage{hyperref}
\usepackage{orcidlink}
\usepackage{tabularx}
\usepackage{lipsum}

\begin{document}

\title{Hybrid neutron stars among mass gap objects are excluded by twin stars at $1.4\,M_\odot$}

\author{Alexander Ayriyan\,\orcidlink{0000-0002-5464-4392}}
\affiliation{Institute of Theoretical Physics, University of Wroclaw, Max Born place 9, 50-204 Wroclaw, Poland}
\affiliation{A. Alikhanyan National Science Laboratory, 
Alikhanyan Brothers street 2, 0036 Yerevan, Armenia}
\email{alexander.ayriyan@gmail.com}

\author{David Blaschke\,\orcidlink{0000-0002-8399-5183}}
\affiliation{Institute of Theoretical Physics, University of Wroclaw, Max Born place 9, 50-204 Wroclaw, Poland}
\affiliation{Helmholtz-Zentrum Dresden-Rossendorf (HZDR),
        Bautzner Landstrasse 400, 01328 Dresden, Germany}
\affiliation{Center for Advanced Systems Understanding (CASUS), Untermarkt 20, 02826 G{\"o}rlitz, Germany}
\email{david.blaschke@uwr.edu.pl}

\author{Marcin Dubaj\,\orcidlink{0009-0005-3589-423X}}
\affiliation{Institute of Independent Studies, ul. Przyjazna 52, 41-706 Ruda \'{S}l\c{a}ska, Poland}
%\email{marcin\_dubaj@wp.pl}

\author{Oleksandr Vitiuk\,\orcidlink{0000-0002-9744-3937}}
\affiliation{Institute of Theoretical Physics, University of Wroclaw, Max Born place 9, 50-204 Wroclaw, Poland}
%\email{oleksandr.vitiuk2@uwr.edu.pl}

\author{Adrian Wojcik\,\orcidlink{0009-0004-4620-8963}}
\affiliation{Institute of Experimental Physics, University of Wroclaw, Max Born place 9, 50-204 Wroclaw, Poland}
%\email{adrian.wojcik@uwr.edu.pl}

\begin{abstract}

We investigate the question whether compact objects in the so-called mass gap ($2.5 < M/M_\odot < 5.0$) can be neutron stars or hybrid stars. Using a generic hybrid star equation of state with a first-order deconfinement transition, we map the allowed parameter space in a Seidov-type diagram and confront it with modern mass-radius constraints. We find that mass-gap hybrid stars require an extremely early onset of deconfinement and very stiff quark matter. The Bayesian analysis, however, favors equations of state with deconfinement at typical neutron-star masses around $1.4\,M_\odot$ with mass-twin stars that, if confirmed, would rule out hybrid stars as candidates for observed mass-gap compact objects.

\end{abstract}

\maketitle

\section{Introduction} 
\label{sec:intro}
Multi-messenger astrophysics provides unprecedented insights into the internal composition of neutron stars (NSs) and properties of strongly interacting matter under extreme conditions~\citep{Koehn:2024set,Ascenzi:2024wws}
by combining signals from gravitational and electromagnetic waves, as well as other cosmic messengers. 
Thus, the mass and radius of NSs are measured by X-ray and radio observations; gravitational wave detections of binary NS and NS-black hole mergers constrain the star's tidal deformability related to the equation of state (EoS) of NS matter~\citep{Abbott_2018,Lattimer:2021emm}.
Gravitational wave signals from merger events of compact objects have led to the detection of objects in the mass gap $2.5 \le M/M_\odot \le 5$, recent examples being: GW170817 
\cite{LIGOScientific:2017vwq} with a total mass of $2.74^{+0.04}_{-0.01}~M_\odot$, GW190814 
\cite{LIGOScientific:2020zkf}
where the lighter component was an object with $2.59^{+0.08}_{-0.09}~M_\odot$ and GW230529 \cite{LIGOScientific:2024elc} 
with a mass of $2.5-4.5~M_\odot$.
The nature of these objects is unclear: are they necessarily black holes or could they be stable neutron stars? 
Since the mass of purely hadronic neutron stars is limited to about $2.1~M_\odot$ \citep{Yamamoto:2015lwa}  because of the EoS softening due to the appearance of heavier baryons such as hyperons and nuclear resonances above 2-3 saturation densities, only hybrid neutron stars could be candidates for non-black-hole mass-gap objects. 
The question arises: Could the hybrid stars reaching the mass gap be from a third family branch \citep{Gerlach:1968zz}?

In 2013, significant progress in the characterization of hybrid neutron star EoS has been made by Alford et al. \cite{Alford:2013aca}, when they identified four regions (A - D) in the Seidov diagram, spanned by the jump of energy density ($\Delta \varepsilon$) at the first-order phase transition against the critical pressure ($P_c$) at its onset energy density ($\varepsilon_c$), where both axes are given in units of this value.
These regions of the EoS parameters would indicate whether the corresponding solutions of the Tolman-Oppenheimer-Volkoff (TOV) equations would result in mass-radius diagrams where stable hybrid star branches are either (A)bsent, (C)onnected or (D)isconnected (third family) with the hadronic star branch.
A special case is the presence of (B)oth a connected and a disconnected hybrid star branch on the mass-radius sequence of a given EoS.

A decisive line on this diagram is the one described by the Seidov criterion \citep{Seidov:1971}
\begin{equation}
    \frac{\Delta \varepsilon}{\varepsilon_c}= 
    \frac{1}{2}+\frac{3}{2}\frac{P_c}{\varepsilon_c}~,
    \label{eq:seidov}
\end{equation}
which divides the plane into region with or without gravitational instability, depending on whether the energy density jump is larger or smaller than the critical value which fulfills Eq. \eqref{eq:seidov}.
Physical properties of neutron stars, such as radii and maximum masses have been shown on this diagram \cite{Alford:2013aca,Alford:2015gna,Ranea-Sandoval:2015ldr,Han:2018mtj}. Recently also the classification of twin stars by
%according to the onset mass 
\cite{Christian:2017jni} has been given in the Seidov diagram \citep{Christian:2023hez}.

\section{Hybrid EoS and M-R diagrams}

In this letter, we shall exploit a generic form of hybrid EoS which has proven suitable for studies of the kind we undertake here. The nuclear matter phase is described in a relativistic density functional of the Walecka type (Walecka+qPauli) for the pressure $P_H(n)$ and the energy density $\varepsilon_H(n)$ as functions of the baryon density $n$, which has an additional repulsive term from quark Pauli blocking between nucleons. It is implemented as a density-dependent shift of the nucleon chemical potentials \citep{Blaschke:2020qrs}. In addition, the nucleon wave-function parameters were tuned to increase the effective-to-bare nucleon mass ratio to $m_N^*/m_N = 0.74$ and to reduce the nuclear matter compressibility to $K = 290.4$ MeV.
The quark matter phase is assumed to have a constant squared sound speed ($c_s^2=dP/d\varepsilon$), which appears to be a good approximation for color superconducting quark matter, where for reasonable parametrizations the range of 
$0.4 < c_s^2 < 0.6$ has been obtained in Ref.~\cite{Contrera:2022tqh}.

\begin{figure}[thb]
\centering
\includegraphics[width=\columnwidth,height=6.5cm]{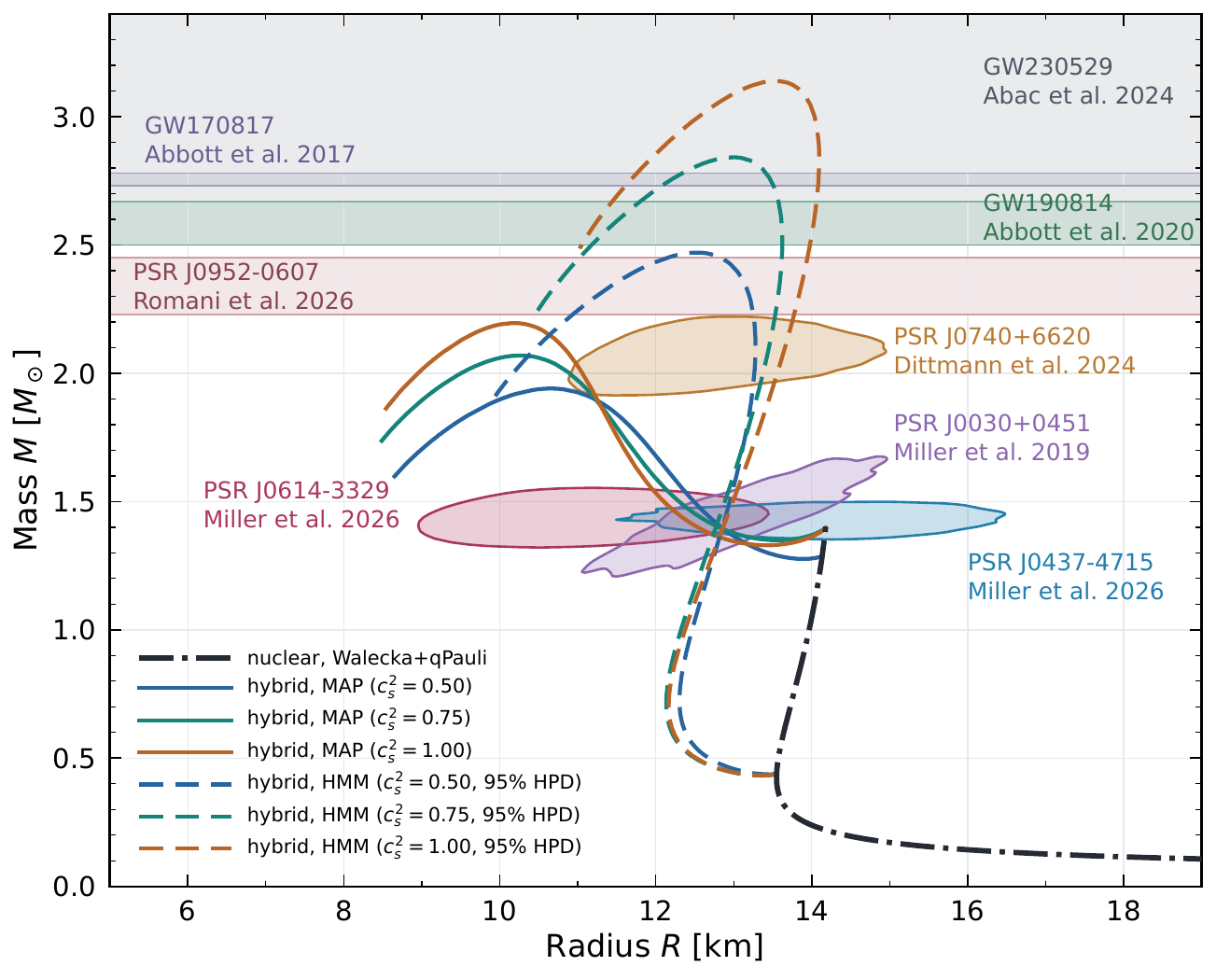}
\caption{Mass--radius diagram for representative parametrizations of the hybrid EoS \eqref{eq:eos}. The black dash-dotted curve shows the purely nuclear EoS \cite{Blaschke:2020qrs}. The other curves correspond to two sets of hybrid star solutions for the EoS \eqref{eq:eos}, where for three values of fixed $c_s^2=0.50, 0.75, 1.00$ each the results of a Bayesian analysis varying ($\varepsilon_c$, $\Delta \varepsilon$) are shown: MAP denotes the EoS with the maximum-a-posteriori probability, while HMM denotes the heaviest maximum-mass models within the parameter region of 95\% credibility.
Colored contours show NICER mass--radius constraints for PSR J0030+0451~\citep{Miller:2019cac},
PSR J0740+6620~\citep{Dittmann:2024mbo},
PSR J0437-4715~\citep{Miller:2025qfq}, and PSR J0614-3329~\citep{Miller:2026vpr}. 
Horizontal bands indicate mass constraints for massive pulsars and the mass-gap objects: GW230529, GW170817, and GW190814 with gray, violet, and green bands,
respectively~\citep{LIGOScientific:2024elc,LIGOScientific:2017vwq,LIGOScientific:2020zkf}. 
The remaining bands show the $1\sigma$ mass constraints for PSR J0348+0432~\citep{Antoniadis:2013pzd} and the black widow pulsar PSR J0952-0607~\citep{Romani:2022jhd, Romani:2025ytn}.
}
\label{fig:m-r}%
\end{figure}
We consider a first-order deconfinement transition that is defined by its onset at the density $n_{\rm onset}$ with the corresponding values of $P_c=P_H(n_{\rm onset})$ and $\varepsilon_c=\varepsilon_H(n_{\rm onset})$ 
with a jump in energy density 
$\Delta \varepsilon=\varepsilon_Q(P_c)-\varepsilon_H(P_c)$, so that the resulting EoS obtains the generic form \cite{Alford:2013aca}
\begin{eqnarray}
   P(\varepsilon)  =
\left\{
	\begin{array}{ll}
		P_H(\varepsilon),  & \varepsilon \leq \varepsilon_c, \\
		P_c, & \varepsilon_c \leq \varepsilon \leq \varepsilon_c + \Delta \varepsilon, %\nonumber
        \\
        P_c +c_s^2(\varepsilon - \varepsilon_c - \Delta \varepsilon), &
        \varepsilon \geq \varepsilon_c + \Delta \varepsilon.
	\end{array}
\right.
\label{eq:eos}
\end{eqnarray}

With this hybrid EoS \eqref{eq:eos}, we can solve the TOV equations and obtain mass-radius diagrams for each set of 
parameters $(\varepsilon_c,\Delta \varepsilon )$,
and for three representative choices of 
$c_s^2=0.50, 0.75, 1.00$. 

Representative examples for TOV solutions are shown in Fig. \ref{fig:m-r} together with modern mass-radius constraints from multi-messenger astronomy. All curves show the mass-twin property: there are two different star configurations with exactly the same gravitational mass, but with different radii. One of them lies at the maximum mass configuration of the hadronic branch (black solid line) while the more compact twin star appears on the so-called "third family" branch of hybrid stars, which is disconnected from the hadronic one by an unstable branch. 
Following these examples, one may hypothesize that 
the two pulsars PSR J0437-4715 \cite{Miller:2025qfq} and PSR J0614-3329 \cite{Miller:2026vpr} 
could be examples of twin stars, but the present level of accuracy for NICER mass and radius measurements does not yet allow a firm conclusion in that respect.

Recent Bayesian analyses have investigated a version of this Seidov diagram in the plane of density jump vs. onset density of the phase transition in units of the nuclear saturation density $n_0=0.15$ fm$^{-3}$, see \cite{Tang:2025xib,Blomqvist:2025cxe}.
This form has the advantage that with densities an intuitive understanding of the reality is associated.
In this letter, we will use an equivalent form of the Seidov relation 
\begin{equation}
    \frac{\Delta n}{n_{\rm onset}}= 
    \frac{1}{2}+\frac{P_c}{P_c+\varepsilon_c}=    \frac{1}{2}+\frac{1}{1+\varepsilon_c/P_c}~,
    \label{eq:seidov-n}
\end{equation}
which is obtained from %Eq. 
\eqref{eq:seidov} by the Gibbs-Duhem relation 
\begin{equation}
\varepsilon=\mu\, n - P    
\label{eq:gibbs-duhem}
\end{equation}
for the states of hadronic and quark matter at the first-order phase transition. They are in mechanical and chemical equilibrium at the same critical pressure $P_c$ and chemical potential $\mu_c$ while the density (energy density) exhibits a jump $\Delta n$ ($\Delta \varepsilon$) for which holds
\begin{equation}
   \Delta \varepsilon= \mu_c\, \Delta n~.
\end{equation}
With this setup of the EoS we can answer our question whether mass gap objects like the ones we mentioned in the Introduction could be not just neutron stars or hybrid stars, but even members of the third family of hybrid stars, thus forming a separate pulsar population.

In addition, we want to give an argument why the existence of a third family and thus of mass-twins of compact stars could be important for pulsar phenomenology.
In a recent investigation, \cite{Chanlaridis:2024rov} 
considered the puzzle of the appearance of millisecond pulsars (MSPs) in eccentric orbits with eccentricities of the order of $e\approx 0.1$ in some low-mass X-ray binary (LMXB) systems (see \cite{Grunthal:2024qfo} for a recent compilation of data for those objects) while generally the orbits of those systems are extremely well circular with
$e<10^{-5}$.
They demonstrated that an accretion-induced phase transition between mass twins on the nucleonic and the hybrid star branches can explain the eccentric orbits when this transition is accompanied with a gravitational mass defect 
$\Delta M/M_{\odot}\approx 0.01$ which induces a kick to the pulsar.
For this scenario to work it is necessary that the transition from hadronic to hybrid star configurations can be completed in a sufficiently small time interval, much smaller than the typical orbital period of $\approx 20 - 30$ days for these objects, see Table 1 of \cite{Grunthal:2024qfo}. 
As it was argued in \cite{Chanlaridis:2024rov}, 
because of angular momentum conservation in this transition, the fast conversion between two configurations is possible only when the final state is on a third family branch.
Conversion to a hybrid star configuration on a connected branch would require a change of angular momentum which takes about $10^4$ years \citep{Glendenning:1997fy}.

In the following section we present the results of a Bayesian inference for the two-parameter EoS \eqref{eq:eos}
with the mass-radius measurements shown in Fig. 
\ref{fig:m-r} using the Seidov diagrams to discuss the question whether hybrid stars can populate the mass gap. 

\begin{figure}
    \centering
    \includegraphics[width=\linewidth]{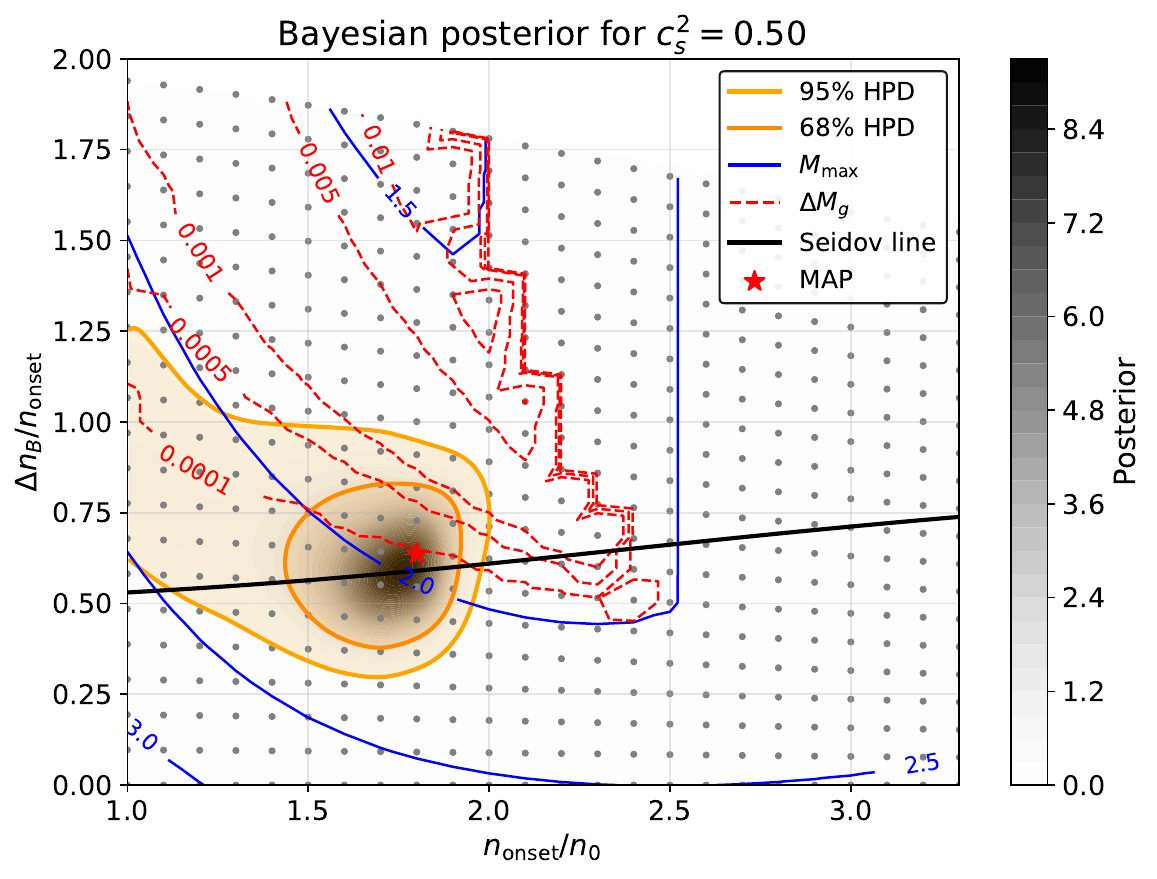}
    \\
    \includegraphics[width=\linewidth]{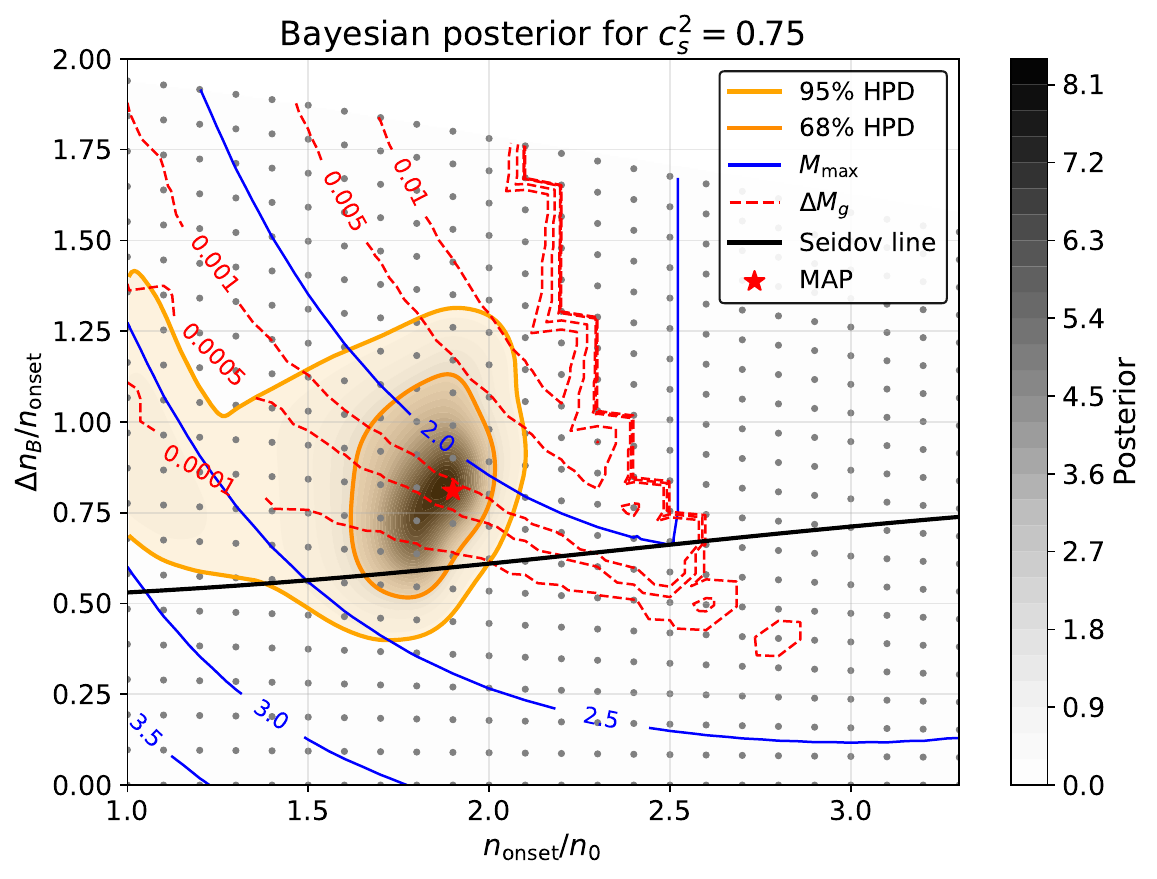}
\\
    \includegraphics[width=\linewidth]{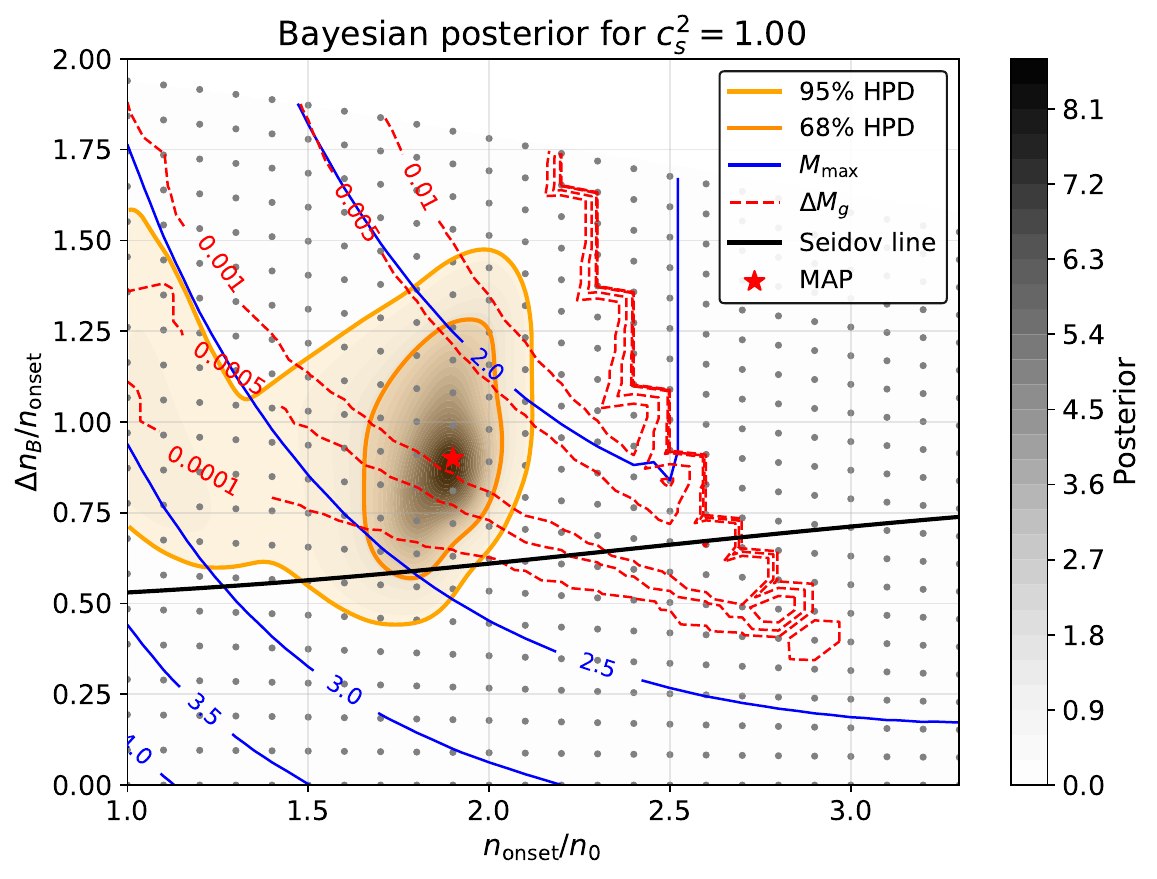}
    \caption{Seidov diagrams, density jump in units of the onset density versus onset density for the hybrid EoS \eqref{eq:eos} and three cases of squared sound speed: 
    $c_s^2=0.50$ (upper panel), 0.75 (middle panel), 1.00 (lower panel).
    The black line denotes the modified Seidov instability criterion \eqref{eq:seidov-n} and blue lines indicate maximum masses of 2.0, 2.5 and 3.0 $M_\odot$, where that for 2.5 $M_\odot$ is the border to the mass gap region. Red lines enclose regions where the mass defect in a transition between twins of equal baryon mass $\Delta M/M_\odot$=0.0001, 0.0005, 0.001, 0.005, 0.01. For details, see text.}
    \label{fig:seidov}
\end{figure}

\section{Results and Discussion}
In Fig. \ref{fig:seidov} we show the Seidov diagrams, i.e. density jump in units of the onset density versus onset density for the hybrid EoS \eqref{eq:eos}, for three cases of squared sound speed: $c_s^2=0.50$ (upper panel), 0.75 (middle panel), 1.00 (lower panel).
The black line denotes the modified Seidov instability criterion \eqref{eq:seidov-n} as a necessary condition for the existence of third family hybrid stars and blue lines indicate maximum masses of 2.0, 2.5 and 3.0 $M_\odot$, where that for 2.5 $M_\odot$ is the border to the mass gap region. Red lines enclose regions where the mass defect in a transition between twins of equal baryon mass $\Delta M/M_\odot$=0.0001, 0.0005, 0.001, 0.005, 0.01.

From these diagrams, one can read off that mass gap objects ($M_{\rm max}\ge 2.5~M_\odot$) 
can be hybrid stars as they populate the lower left corner of all three Seidov diagrams. The highest maximum masses are obtained for crossover type transitions ($\Delta n=0$), lowest possible transition densities and maximum stiffness of quark matter. As it has been recently reported in a paper \cite{Blaschke:2026mpq} revisiting the Rhoades-Ruffini bound \citep{Rhoades:1974fn}, maximum masses of hybrid stars may exceed $4~M_\odot$, see the lower-left corner of the bottom panel in Fig. \ref{fig:seidov}, if the onset of deconfinement could be around the nuclear saturation density and the speed of sound could reach the speed of light. 
However, in order to find cases of third-family objects among the high-mass hybrid stars, the density jump at the deconfinement transition has to exceed the value determined by the modified Seidov criterion, which is shown by the black line in Fig. \ref{fig:seidov}. 

For the typical value of the squared sound speed in color superconducting quark matter models \citep{Contrera:2022tqh}, $c_s^2=0.5$, one can find mass-gap objects for all onset densities below $2.5\, n_0$.
But third family members among them are just a marginal case at the left border of the Seidov diagram, for $n_{\rm onset}=n_0$ and $\Delta n=0.6\,n_0$. 

Following the arguments of \cite{Hippert:2024hum} on the basis of relativistic hydrodynamic transport, the maximum possible squared sound speed in dense nuclear matter is $c_s^2\le 0.781$, well represented by the case of the middle panel in Fig. \ref{fig:seidov}.
There, third-family members are found among the mass-gap objects for all onset densities 
fulfilling $n_{\rm onset}\le 1.5\, n_0$.
However, the transition from hadronic star configurations to their third-family mass twins involves mass defects of only $\sim 10^{-4}\, M_\odot$ which are too small to generate a sufficient kick that would explain the observed eccentricities of some MSPs in LMXBs. In order to be relevant for explaining the eccentric MSP puzzle, the onset mass for the deconfinement transition should exceed 
$\sim 1.2\, M_\odot$, which concerns onset densities $n_{\rm onset}\gtrsim 1.7\, n_0$. 
The interesting case of $M_{\rm onset}\sim 1.4\, M_\odot$ (see Fig. \ref{fig:m-r}) would imply $n_{\rm onset}\sim 1.9\, n_0$. 
Figure~\ref{fig:seidov} shows that for these onset densities, mass gap objects are below the Seidov line and thus cannot be members of a third family branch.

So far we were focused on discussing the theoretical possibility that mass gap objects could be members of the third family of hybrid stars and have identified the conditions under which this could be the case. 
We should, however, not ignore the existing observational constraints on masses and radii of neutron stars which need to be simultaneously fulfilled. In order to take into account these $M$-$R$ constraints depicted in Fig. \ref{fig:m-r}, we apply the Bayesian analysis scheme described in \cite{Ayriyan:2024zfw}, but with the quark matter EoS parameters $(n_{\rm onset},\Delta n)$ instead of the Lagrangian model parameters for the coupling strengths in the vector meson and diquark interaction channels $(\eta_V,\eta_D)$ of that model approach.
The results are shown as a grayscale heatmap in Fig. \ref{fig:seidov}. 
We want to focus our discussion on the top and middle panels since the case of the bottom panel ($c_s^2=1.00$) cannot be considered a realistic one. 
There are solutions with an early onset of deconfinement, $1.0 \le n_{\rm onset}/n_0\le 1.5$, 
corresponding to onset masses below $0.5\,M_\odot$, which have maximum masses can reach also into the mass gap region, see Fig. \ref{fig:m-r} for the corresponding $M(R)$ lines.
Such solutions have recently been identified as favorable ones within a Bayesian analysis performed in \cite{Ayriyan:2025rub}.

The other solutions with high posterior probability correspond to onset masses with twin star solutions around $1.4\,M_\odot$, but for these sequences maximum masses just reach beyond $2\,M_\odot$ and cannot reach the mass gap region.
This latter solution class could support the speculation that PSR J0614-3329 and PSR J0437-4715 (possibly including PSR J0030+0451) could be mass twin stars, see Fig.  \ref{fig:m-r}. This EoS would then support the scenario of an accretion induced deconfinement transition as a solution of the puzzle of eccentric MSP orbits 
\citep{Chanlaridis:2024rov}.  

\section{Conclusions}
In this letter we have shown that for a generic two-parametric hybrid neutron star EoS in dependence on the onset density and density jump at the deconfinement transition there exist parameter sets for which the maximum mass is in the mass gap region,  $M_{\rm max} > 2.5\, M_\odot$.
These sequences belong to an early onset of deconfinement at $M_{\rm onset}\lesssim 0.7\, M_\odot$ with onset densities 
$1.0 \le n_{\rm onset}/n_0\le 1.2$.
In order to justify that such low onset densities may be plausible, we refer to arguments from the occupation of fermionic phase space in cold, degenerate (quarkyonic) matter, where due to quark exchange effects among nucleons (quark Pauli blocking, see \cite{Blaschke:2020qrs}) just above the nuclear saturation density for 
$n_T=0.17\, {\rm fm}^{-3}=1.13\,n_0$ 
a transformation of nuclear matter to quarkyonic matter sets in \citep{Koch:2024qnz,McLerran:2024rvk}. 

Should, however, the conjecture be confirmed that 
the pulsars PSR J0614-3329 and PSR J0437-4715 are 
mass twin stars at $1.4\,M_\odot$, then the maximum mass for such a hybrid EoS shall be limited to 
$M_{\rm max} < 2.2\, M_\odot$ so that the detected 
mass gap objects cannot be hybrid neutron stars but rather must be black holes.

\section*{Acknowledgments}
We thank Cole Miller for bringing Refs. \cite{Miller:2026vpr,Romani:2025ytn} to our attention and for his encouragement to complete this work.
This research is part of the project No. 2021/43/P/ST2/03319 co-funded by the National Science Centre and the European Union’s Horizon 2020 research and innovation programme under the Marie Skłodowska-Curie grant agreement no. 945339. Additionally, this research was funded in whole or in part by the National Science Centre, Poland, under Grant No. 2022/45/N/ST2/02391. 

For the purpose of Open Access, the author has applied a CC-BY public copyright licence to any Author Accepted Manuscript (AAM) version arising from this submission.

\section*{Data Availability}

The numerical data supporting the results and figures presented in this work,
including the equation-of-state, mass--radius, and Bayesian inference tables,
are publicly available on Zenodo \citep{Ayriyan:2026data}.

\end{document}